%% file: HIG-11-029_temp.tex
\pdfoutput=1

\documentclass[11pt,twoside,a4paper,cmspaper,final,collab]{cms-tdr}
\def\svnVersion{267487}\def\cmsCernNo{2013-037}\def\cmsCernDate{\today}\def\cmsMessage{Published in Physics Letters B as \href{http://dx.doi.org/10.1016/j.physletb.2012.05.028}{\doi{10.1016/j.physletb.2012.05.028.}}}

\begin{document}\cmsNoteHeader{HIG-11-029}

\hyphenation{had-ron-i-za-tion}
\hyphenation{cal-or-i-me-ter}
\hyphenation{de-vices}

\RCS$Revision: 116407 $
\RCS$HeadURL: svn+ssh://svn.cern.ch/reps/tdr2/papers/HIG-11-029/trunk/HIG-11-029.tex $
\RCS$Id: HIG-11-029.tex 116407 2012-04-16 19:42:39Z klute $
\newcommand{\xsecbr}{\sigma_\phi\cdot B_{\Pgt\Pgt}}
\newcommand{\xsecbrH}{\sigma_H\cdot B_{\Pgt\Pgt}}
\newcommand{\MT}{m_\mathrm{T}}
\newcommand{\Mvis}{m_\text{vis}}
\newcommand{\Mfit}{m_{\Pgt\Pgt}}
\providecommand{\mH}{\ensuremath{m_{\PH}}\xspace}
\newlength\cmsFigWidth
\ifthenelse{\boolean{cms@external}}{\setlength\cmsFigWidth{0.95\columnwidth}}{\setlength\cmsFigWidth{0.8\textwidth}}
\ifthenelse{\boolean{cms@external}}{\providecommand{\cmsLeft}{top}}{\providecommand{\cmsLeft}{left}}
\ifthenelse{\boolean{cms@external}}{\providecommand{\cmsRight}{bottom}}{\providecommand{\cmsRight}{right}}
\cmsNoteHeader{HIG-11-029} 
\title{Search for neutral Higgs bosons decaying to tau pairs in $\Pp\Pp$ collisions at $\sqrt{s}=7\TeV$}

\date{\today}

\abstract{
A search for neutral Higgs bosons decaying to tau pairs at a center-of-mass energy of 7\TeV is performed using a dataset corresponding to an integrated luminosity of 4.6\fbinv recorded  by the CMS experiment at the LHC. The search is sensitive to both the standard model Higgs boson and to the neutral Higgs bosons predicted by the minimal supersymmetric extension of the standard model (MSSM). No excess of events is observed in the tau-pair invariant-mass spectrum. For a standard model Higgs boson in the mass range of 110--145\GeV upper limits at 95\% confidence level (CL) on the production cross section are determined. We exclude a Higgs boson with $\mH=115\GeV$ with a production cross section 3.2 times of that predicted by the standard model. In the MSSM, upper limits on the neutral Higgs boson production cross section times branching fraction to tau pairs, as a function of the pseudoscalar Higgs boson mass, $m_\mathrm{A}$, sets stringent new bounds in the parameter space, excluding at 95\% CL values of $\tan\beta$ as low as 7.1 at $m_\mathrm{A}=$160\GeV in the $m_\mathrm{h}^{\rm max}$ benchmark scenario.}

\hypersetup{%
pdfauthor={CMS Collaboration},%
pdftitle={Search for neutral Higgs bosons decaying to tau pairs in pp collisions at sqrt(s)=7 TeV},%
pdfsubject={CMS},%
pdfkeywords={CMS, physics, Higgs}}

\maketitle 

\section{Introduction}
An important goal of the LHC physics program is to ascertain the mechanism of electroweak symmetry breaking, through which the \PW\ and \cPZ\ bosons attain mass, while the photon remains massless. In the standard model (SM)~\cite{SM1,SM2,SM3}, this is achieved via the Higgs mechanism~\cite{Englert:1964et,Higgs:1964ia,Higgs:1964pj,Guralnik:1964eu,Higgs:1966ev,Kibble:1967sv}, which also predicts the existence of a scalar Higgs boson. However, this particle has not yet been observed by experiments. Moreover, the mass of the Higgs boson is quadratically divergent at high energies~\cite{Witten-hierarchy}. Supersymmetry~\cite{Martin97} is a well known extension to the SM which allows the cancellation of this divergence.

The minimal supersymmetric standard model (MSSM) contains two Higgs doublets, giving rise to five physical states: a light neutral CP-even state (h), a heavy neutral CP-even state (\PH), a neutral CP-odd state (A), and a pair of charged states ($\PH^\pm$)~\cite{fayet1, fayet2, glashow-2hdm, deshpande}. The mass relations between these particles depend on the MSSM parameter $\tan\beta$, the ratio of the Higgs fields vacuum expectation values. We focus on the $m_\mathrm{h}^{\rm max}$~\cite{MHMAX-Carena-2002,MHMAX-Carena} benchmark scenario in which $M_{\rm SUSY}$ = 1~TeV; $X_t$ =2$M_{\rm SUSY}$; $\Pgm$ =~200\GeV; $M_{\tilde{g}}$ = 800\GeV; $M_2$ = 200\GeV; and $A_b = A_t$. Here, $M_{\rm SUSY}$ denotes the common soft-SUSY-breaking squark mass of the third generation; $X_t = A_t - \Pgm/\tan\beta$ is the stop mixing parameter; $A_t$ and $A_b$ are the stop and sbottom trilinear couplings, respectively; $\Pgm$ the Higgsino mass parameter; $M_{\tilde{g}}$ the gluino mass; and $M_2$ is the SU(2)-gaugino mass parameter. The value of $M_1$ is fixed via the unification relation $M_1 =(5/3)M_2\sin\theta_{\rm W}/\cos\theta_{\rm W}$. In this scenario for values of $\tan\beta \gtrsim 15$, if $m_\mathrm{A} \lesssim$ 130\GeV the masses of the h and A are almost degenerate, while the mass of the \PH is around 130\GeV. Conversely, if $m_\mathrm{A} \gtrsim$ 130\GeV, the masses of the A and H are almost degenerate, while the mass of the h remains near 130\GeV. This will thus always lead to one neutral Higgs boson at 130\GeV and two neutral Higgs bosons with almost degenerate mass of $m_\mathrm{A}$. 

Direct searches for the SM Higgs boson at the Large Electron-Positron Collider (LEP) set a limit on
the mass $m_{\PH} > 114.4\GeV$ at 95\% confidence level (CL)~\cite{Barate:2003sz}.
The Tevatron collider experiments exclude the SM Higgs boson in the mass range 162--166\GeV~\cite{TEVHIGGS_2010},
and the ATLAS experiment in the mass ranges 112.9--115.5, 131--238, and 251--466\GeV~\cite{AtlasCombination}.
Precision electroweak data constrain the mass of the SM Higgs boson to be less than 158\GeV~\cite{EWK}.
Direct searches for neutral MSSM Higgs bosons have been reported by LEP, the Tevatron, and both LHC experiments, and set limits on the MSSM parameter space in the $\tan\beta$--m$_\mathrm{A}$ plane~\cite{LEP2-MSSM, CDF-MSSM, D0-MSSM, ATLAS-MSSM, CMS-PAPERS-HIG-10-002}.

This Letter reports a search for the SM and the neutral MSSM Higgs
bosons using final states with tau pairs in proton-proton collisions at $\sqrt{s}=7$\TeV at the LHC.
We use a data sample collected in 2011 corresponding to an
integrated luminosity of 4.6\fbinv recorded by the Compact Muon Solenoid (CMS)~\cite{CMS-JINST}
experiment. Three independent tau pair final states where one or both taus
decay leptonically are studied: $\Pe\Pgt_h$+X, $\Pgm\Pgt_h$+X, and
$\Pe\Pgm$+X, where we use the symbol $\Pgt_h$ to indicate a
reconstructed hadronic decay of a tau.

In the case of the SM Higgs boson, the gluon-fusion production mechanism has the largest cross section. However, in the mass region of interest, background from Drell--Yan production of tau pairs overwhelms the expected Higgs boson signal. This search therefore relies upon the signature of Higgs bosons produced via vector boson fusion (VBF) or in association with a high-\pt jet. In the former case, the distinct topology of two jets with a large rapidity separation greatly reduces the background. In the latter, requiring a high-\pt jet both suppresses background, and improves the measurement of the tau-pair invariant mass.

In the MSSM case, two main production processes contribute to
$\Pp\Pp\to\phi$+X, where $\phi=$~h, H, or A: gluon fusion through a
\cPqb-quark loop and direct $\cPqb\cPaqb$ annihilation from the
\cPqb-quark content of the beam protons.  In the latter case, there
is a significant probability that a b-quark jet is produced
centrally in association with the Higgs boson due to the enhanced
$\cPqb\cPaqb\phi$ coupling.  Requiring a b-quark jet increases
the sensitivity of the search by reducing the $\cPZ + $jets background.

\section{CMS detector}

The CMS detector is described in detail elsewhere~\cite{CMS-JINST}. The central feature of the CMS apparatus is a superconducting solenoid of 6~m internal diameter, providing a magnetic field of 3.8\unit{T}. Within the solenoid are the silicon pixel and strip tracker, which cover a pseudorapidity region of $|\eta| < 2.5$. Here, the pseudorapidity is defined as $\eta=-\ln{(\tan{\theta/2})}$, where $\theta$ is the polar angle of the trajectory of the particle with respect to the direction of the counterclockwise beam. The lead tungstate crystal electromagnetic calorimeter and the brass-scintillator hadron calorimeter surround the tracking volume and cover $|\eta| < 3$.  In addition to the barrel and endcap detectors, CMS has extensive forward calorimetry which extends the coverage to $|\eta| < 5$. Muons are measured in gas-ionization detectors embedded in the steel return yoke, with a coverage of $|\eta| < 2.4$.

\section{Trigger and event selection}

The analysis makes use of the three independent tau-pair final states, $\Pe\Pgt_h$+X, $\Pgm\Pgt_h$+X, and $\Pe\Pgm$+X.
In all three channels, there is substantial background, both from processes with similar experimental signatures,
and from unrelated hadronic activity in the detector.

The trigger selection required a combination of electron, muon and tau trigger objects~\cite{CMS-PAS-EGM-10-004,CMS-PAS-MUO-10-002,CMS-EWK-TAU}.
The identification criteria and $\pt$
thresholds of these objects were progressively
tightened as the LHC instantaneous luminosity increased over the data-taking period.

A particle-flow algorithm~\cite{CMS-PAS-PFT-09-001,CMS-PAS-PFT-10-002,CMS-PAS-PFT-10-003} is
used to combine information from all CMS subdetectors to identify
and reconstruct individual particles in the event, namely muons,
electrons, photons, and charged and neutral hadrons.
From the resulting particle list jets, hadronically-decaying taus, and missing transverse energy (\MET), defined as the negative of the vector sum of the transverse momenta, are reconstructed. The jets are identified using the anti-$k_T$ jet algorithm~\cite{Cacciari:fastjet1,Cacciari:fastjet2} with a distance parameter of $R=0.5$. Hadronically-decaying taus are reconstructed using the hadron plus strips (HPS) algorithm, which considers candidates with one or three charged pions and up to two neutral pions~\cite{CMS-PAS-TAU-11-001}.

For the $\Pe\Pgt_h$+X and $\Pgm\Pgt_h$+X final states, in the region $|\eta| < 2.1$, we select events with an electron of $\pt >$~20\GeV or a muon of $\pt >17\GeV$, together with an oppositely charged $\Pgt_h$ of $\pt > 20\GeV$ within the range $|\eta| < 2.3$. For the $\Pe\Pgm$+X final state, we select events with an electron of $|\eta|<2.3$ and an oppositely charged muon of $|\eta|<2.1$, requiring $\pt >$ 20\GeV for the highest-\pt lepton and $\pt >10\GeV$ for the next-to-highest-\pt lepton. For the $\Pe\Pgt_h$+X and $\Pgm\Pgt_h$+X final states, we reject events with more than one electron or more than one muon of $\pt >15\GeV$.

Taus from Higgs boson decays are typically isolated from the rest of the event activity, in contrast to background from jets, which are typically immersed in considerable hadronic activity. For each lepton candidate ($\Pe$, $\Pgm$, or $\Pgt_h$), a cone is constructed around the lepton direction at the event vertex.
An isolation variable is constructed from the scalar sum of the transverse energy of all reconstructed
particles contained within the cone, excluding the contribution from the lepton candidate itself.

In 2011, an average of ten proton-proton interactions occurred per LHC bunch crossing, making the assignment of the vertex of the hard-scattering process non-trivial. For each reconstructed collision vertex, the sum of the $\pt^2$ of all tracks associated to the vertex
is computed. The vertex for which this quantity is the largest is assumed to correspond to the hard-scattering
process, and is referred to as the primary vertex.
A correction is applied to the isolation variable to account for effects of
additional interactions. For charged particles, only those associated with the primary vertex are considered in the isolation variable.
For neutral particles, a correction is applied
by subtracting the energy deposited in the isolation cone by charged particles not
associated with the primary vertex, multiplied by a factor of 0.5. This factor corresponds approximately
to the ratio of neutral to charged hadron production in the hadronization process of pile-up interactions.
An $\eta$, \pt, and lepton-flavor dependent threshold on the isolation variable of
 less than roughly 10\% of the candidate \pt is applied.

To correct for the contribution to the jet energy due to pile-up, a median energy density ($\rho$) is
determined event by event. The pile-up contribution to the jet energy is estimated as the product
of $\rho$ and the area of the jet and subsequently subtracted from the jet transverse energy~\cite{Cacciari:subtraction}.
In the fiducial region for jets of $|\eta| < 4.7$, jet energy corrections are also applied as a function of the jet \ET and $\eta$~\cite{cmsJEC}.

In this analysis, due to the small mass
of the tau and the large transverse momentum, the
neutrinos produced in the decay tend to be produced nearly collinear with the visible products.
Conversely, in $\PW+$jets events, one of the main backgrounds, the high mass
of the $\PW$ results in a neutrino approximately opposite to the lepton in the transverse plane, while
 a jet is misidentified as a tau.
In the $\Pe\Pgt_h$+X and $\Pgm\Pgt_h$+X channels of the SM Higgs boson search, which focuses on lower masses (less than 145\GeV), we therefore require the transverse mass
\begin{equation}
\MT = \sqrt{2 \pt \MET (1-\cos(\Delta\phi))}
\end{equation}
to be less than 40\GeV, where \pt is the lepton transverse momentum,
and $\Delta\phi$ is the difference in $\phi$ of the lepton and \MET vector.

In the MSSM search channels and in the $\Pe\Pgm$+X SM search channel, we use a
discriminator formed by considering the bisector of the directions of
the visible tau decay products transverse to the beam direction,
denoted as the $\zeta$ axis~\cite{CRISTOBAL}.  From the
projections of the visible decay product momenta and the $\MET$
vector onto the $\zeta$ axis, two values are calculated:
\begin{equation}
 P_{\zeta} = p_\mathrm{T,1} \cdot \zeta + p_\mathrm{T,2} \cdot \zeta+\MET \cdot \zeta,
\end{equation}
\begin{equation}
 P_{\zeta}^{\text{vis}} = p_\mathrm{T,1} \cdot \zeta + p_\mathrm{T,2} \cdot \zeta.
\end{equation}
where the indices $p_\mathrm{T, 1}$ and $p_\mathrm{T, 2}$ indicate the transverse momentum
of two reconstructed leptons.
For the $\Pe\Pgt_h$+X and $\Pgm\Pgt_h$+X channels in the MSSM search we require
$P_\zeta- 0.5 \cdot P_\zeta^{\text{vis}} > -20$\GeV and for the $\Pe\Pgm$+X channel
we require $P_\zeta- 0.85 \cdot P_\zeta^{\mathrm{vis}} > -25$\GeV.

To further enhance the sensitivity of the search for Higgs bosons both
in the MSSM and in the SM, we split the sample of selected events into
several mutually exclusive categories based on the jet multiplicity and b-jet content.

In the MSSM case, there is a large probability for having a b-tagged jet in the central
region. We use an algorithm based on the impact parameter of the tracks
associated to the event vertex to identify b-tagged jets~\cite{BTV-10-001}.
The MSSM search has two categories:
\begin{itemize}
\item{\textbf{\emph{b-Tag category:}}} We require at most one jet with $\pt > 30$\GeV and
at least one b-tagged jet with $\pt > 20\GeV$.
\item{\textbf{\emph{Non b-Tag category:}}} We require at most one jet with $\pt > 30$\GeV and
no b-tagged jet with $\pt > 20$\GeV.
\end{itemize}

The SM search has three categories:
\begin{itemize}
\item{\textbf{\emph{VBF category:}}} We require at least two jets with $\pt > 30$\GeV,
$|\Delta\eta_{jj}|>4.0$, $\eta_1\cdot\eta_2<0$,
and a dijet invariant mass $m_{jj} >$ 400\GeV, with no other jet with
$\pt > 30$\GeV in the rapidity region between the two jets.
\item{\textbf{\emph{Boosted category:}}} We require one jet with
$\pt >150$\GeV, and, in the $\Pe\Pgm$ channel, no b-tagged jet with $\pt > 20$\GeV.
\item{\textbf{\emph{0/1 Jet category:}}} We require no more than one jet with
$\pt > 30$\GeV, and if such a jet is present, it must have $\pt <150\GeV$.
\end{itemize}

The observed number of events for each category, as well as the expected number of events from various background processes
are shown in Tables~\ref{tab:cutflow-etau}--\ref{tab:cutflow-emu} together with expected signal yields and efficiencies.
The largest source of events selected with these
requirements is $\cPZ\to\Pgt\Pgt$ decays.
We estimate the contribution from this process using an observed sample of $\cPZ\to\Pgm\Pgm$ events, where the reconstructed muons are replaced by the reconstructed particles from simulated tau decays, a procedure called 'embedding'.
The normalization for this process is determined from the
measurement of the $\cPZ\to\Pe\Pe$ and $\cPZ\to\Pgm\Pgm$ cross section~\cite{CMS-EWK-WZ}.

Another significant source of background is multijet events in which there is one jet misidentified as an isolated electron or muon, and a second jet misidentified as $\Pgt_h$. $\PW$+jets events
in which there is a jet misidentified as a $\Pgt_h$ are also a source of background.  The rates for
these processes are estimated using the number of observed same-charge
tau pair events, and from events with large transverse mass, respectively.
Other background processes include $\cPqt\cPaqt$ production
and $\cPZ\to \Pe\Pe/\Pgm\Pgm$ events, particularly in the $\Pe\Pgt_h$+X
channel due to the 2--3\% probability for electrons to be
misidentified as $\Pgt_h$~\cite{CMS-PAS-TAU-11-001}.
The small background from $\PW +$jets and multijet events for the $\Pe\Pgm$
channel where jets are misidentified as isolated leptons is derived by measuring the number of events with one good lepton and a second
which passes relaxed selection criteria, but fails the nominal lepton selection.
This sample is extrapolated to the signal region using the efficiencies
for such loose lepton candidates to pass the nominal lepton selection. These efficiencies
are measured in data using multijet events. Backgrounds from \ttbar  and di-boson production
are estimated from simulation 
using the \MADGRAPH~\cite{Alwall:2007st} event generator to simulate the shapes for \ttbar events and \PYTHIA 6.424~\cite{Pythia} to simulate the shapes for di-boson events. 
The event yields are
normalized to the inclusive cross sections: $\sigma_{\ttbar} = 164.4 \pm 14.3\unit{pb}$ and $\sigma_{\PW\PW} = 55.3 \pm 8.3$\unit{pb}
as measured with an analysis similar to that described in~\cite{CMS-TOP-XSEC, CMS-EWK-DIW} using a larger data sample.

To model the MSSM and SM Higgs boson signals the event generators \PYTHIA and \textsc{powheg}~\cite{POWHEG2} are used, respectively.
The \TAUOLA~\cite{TAUOLA} package is used for tau
decays in all cases. Additional next-to-next-to-leading order (NNLO) K-factors from \textsc{FeHiPro}~\cite{FeHiPro1,FeHiPro2} are  applied to the Higgs boson \pt spectrum from Higgs boson events produced via gluon fusion.

The presence of pile-up is incorporated by simulating additional interactions and then reweighting the
simulated events to match the distribution of additional
interactions observed in data. The events in the embedded $\cPZ\to\Pgt\Pgt$ sample
and in other background samples obtained from data contain
the correct distribution of pile-up interactions.  The missing transvere energy response from simulation is corrected using a prescription, based on data, developed for inclusive \PW\ and \cPZ\ cross section measurements~\cite{CMS-EWK-WZ}, where \cPZ\ bosons are reconstructed in the dimuon channel, and the missing transvere energy scale and resolution calibrated as a function of the \cPZ\ boson transverse momentum.

\begin{table*}[!htbp]
\begin{center}
\caption{Numbers of expected and observed events
  in the event categories as described in the text for the
  $\Pe\Pgt_h$+X channel. Also given are the expected signal yields and efficiencies for a MSSM
  Higgs boson with \mbox{$m_\mathrm{A}=120$\GeV} and $\tan\beta$ = 10, and for a SM Higgs boson with
  \mbox{$\mH=120$\GeV}. Combined statistical and systematic uncertainties on each estimate are reported. 
  For the yield estimates for the Higgs signal the production cross sections for h and A, which have almost 
  degenerate masses, are taken into account.
 The quoted efficiencies do not include the branching fraction into $\Pgt\Pgt$.}
 \begin{tabular}{|l|r@{$ \,\,\pm\,\, $}l|r@{$\,\,\pm\,\,$}l|r@{$\,\,\pm\,\,$}l|r@{$\,\,\pm\,\,$}l|r@{$\,\,\pm\,\,$}l|}
 \cline{2-11}
\multicolumn{1}{c|}{ } &  \multicolumn{6}{c|}{SM}  & \multicolumn{4}{c|}{MSSM} \\
\hline
Process & \multicolumn{2}{c|}{\emph{0/1-Jet}}  &   \multicolumn{2}{c|}{\emph{Boosted}} &   \multicolumn{2}{c|}{\emph{VBF}}  &   \multicolumn{2}{c|}{\emph{Non b-Tag}}  &   \multicolumn{2}{c|}{\emph{b-Tag}} \\
\hline
$\cPZ\rightarrow \Pgt\Pgt$ & 13438 & 977 & 190 & 14 & 19 & 1 & 14259 & 1037 & 135 & 9 \\
Multijets & 6365 & 299 & 27 & 3 & 15 & 2 & 6404 & 301 & 100 & 7 \\
\PW+jets & 2983 & 216 & 62 & 4 & 4.2 & 0.4 & 5432 & 377 & 39 & 3 \\
$\cPZ\rightarrow ll$ & 5170 & 464 & 28 & 4 & 5 & 1 & 6146 & 502 & 28 & 4 \\
\ttbar & 63 & 7 & 42 & 6 & 2 & 1 & 47 & 7 & 75 & 11 \\
Dibosons & 68 & 21 & 5 & 2 & 0.1 & 0.1 & 105 & 22 & 1 & 1 \\
\hline
\hline
Total Background & 28087 & 1142 & 354 & 17 & 45 & 2.9 & 32392 & 1249 & 378 & 17 \\
\hline
$\PH\rightarrow\tau\tau$ &53 & 9 &2.7 & 0.6 &2.0 & 0.2 &279 & 29 &26 & 4 \\
\hline
Data & \multicolumn{2}{c|}{27727} &\multicolumn{2}{c|}{318} &\multicolumn{2}{c|}{43} &\multicolumn{2}{c|}{32051} &\multicolumn{2}{c|}{391}\\
\hline
\multicolumn{6}{c}{ } \\
\multicolumn{2}{l}{Signal Efficiency} &  \multicolumn{4}{c}{ } \\
\hline
gg$\rightarrow \phi$\vphantom{\LARGE h} &  \multicolumn{2}{c|}{-} &     \multicolumn{2}{c|}{-} &     \multicolumn{2}{c|}{-} & \multicolumn{2}{c|}{1.0$\cdot 10^{-2}$} & \multicolumn{2}{c|}{9.0$\cdot 10^{-5}$}  \\
gg$\rightarrow\cPqb\cPqb\phi$ &  \multicolumn{2}{c|}{-} &     \multicolumn{2}{c|}{-} &    \multicolumn{2}{c|}{-} & \multicolumn{2}{c|}{1.1$\cdot 10^{-2}$} & \multicolumn{2}{c|}{1.5$\cdot 10^{-3}$}      \\
\hline
gg$\rightarrow\PH$ \vphantom{\LARGE h} &   \multicolumn{2}{c|}{9.1$\cdot 10^{-3}$} &  \multicolumn{2}{c|}{2.9$\cdot 10^{-4}$} &  \multicolumn{2}{c|}{2.9$\cdot 10^{-5}$} &    \multicolumn{2}{c|}{ -} &     \multicolumn{2}{c|}{-}  \\
qq$\rightarrow\cPq\cPq\PH$ &  \multicolumn{2}{c|}{5.2$\cdot 10^{-3}$} &  \multicolumn{2}{c|}{1.6$\cdot 10^{-3}$} &  \multicolumn{2}{c|}{3.3$\cdot 10^{-3}$} &     \multicolumn{2}{c|}{-} &     \multicolumn{2}{c|}{-}       \\
qq$\rightarrow\PH\ttbar$ or VH&  \multicolumn{2}{c|}{7.8$\cdot 10^{-3}$} &  \multicolumn{2}{c|}{2.2$\cdot 10^{-3}$} &  \multicolumn{2}{c|}{2.8$\cdot 10^{-5}$} &     \multicolumn{2}{c|}{-} &     \multicolumn{2}{c|}{-}       \\
\hline
\end{tabular}
\label{tab:cutflow-etau}
\end{center}
\end{table*}

\begin{table*}[!htbp]
\begin{center}
\caption{Numbers of expected and observed events
  in the event categories as described in the text for the
  $\Pgm\Pgt_h$+X channel. Also given are the expected signal yields and efficiencies for a MSSM
  Higgs boson with \mbox{$m_\mathrm{A}=120$\GeV} and $\tan\beta$ = 10, and for a SM Higgs boson with
  \mbox{$\mH=120$\GeV}.  Combined statistical and systematic uncertainties on each estimate are reported.
  For the yield estimates for the Higgs signal the production cross sections for h and A, which have almost 
  degenerate masses, are taken into account.
  The quoted efficiencies do not include the branching fraction into $\Pgt\Pgt$.}
\begin{tabular}{|l|r@{$ \,\,\pm\,\, $}l|r@{$\,\,\pm\,\,$}l|r@{$\,\,\pm\,\,$}l|r@{$\,\,\pm\,\,$}l|r@{$\,\,\pm\,\,$}l|}
\cline{2-11}
\multicolumn{1}{c}{ } &  \multicolumn{6}{|c|}{SM}  & \multicolumn{4}{c|}{MSSM} \\
\hline
Process & \multicolumn{2}{c|}{\emph{0/1-Jet}}  &   \multicolumn{2}{c|}{\emph{Boosted}} &   \multicolumn{2}{c|}{\emph{VBF}}  &   \multicolumn{2}{c|}{\emph{Non b-Tag}}  &   \multicolumn{2}{c|}{\emph{b-Tag}} \\
\hline
Z$\rightarrow \tau\tau$ & 28955 & 2054 & 295 & 22 & 36 & 2 & 29795 & 2114 & 259 & 18 \\
Multijets & 7841 & 141 & 36 & 2 & 23 & 2 & 6387 & 115 & 160 & 9 \\
W+jets & 5827 & 392 & 65 & 4 & 9 & 1 & 9563 & 628 & 110 & 9 \\
Z$\rightarrow ll$ & 777 & 70 & 5 & 1 & 1.0 & 0.2 & 924 & 115 & 3 & 1 \\
$\ttbar$ & 147 & 15 & 94 & 12 & 4 & 1 & 101 & 15 & 145 & 20 \\
Dibosons & 178 & 55 & 9 & 4 & 0.4 & 0.4 & 217 & 46 & 5 & 2 \\
\hline
\hline
Total Background & 43725 & 2097 & 504 & 26 & 73 & 3.9 & 46987 & 2211 & 681 & 30 \\
\hline
H$\rightarrow\tau\tau$ &96 & 17 &3.9 & 0.8 &3.0 & 0.5 &502 & 52 &45 & 6 \\
\hline
Data & \multicolumn{2}{c|}{43612} &\multicolumn{2}{c|}{500} &\multicolumn{2}{c|}{76} &\multicolumn{2}{c|}{47178} &\multicolumn{2}{c|}{680}\\
\hline
\multicolumn{6}{c}{ } \\
\multicolumn{2}{l}{Signal Efficiency} &  \multicolumn{4}{c}{ } \\
\hline
gg$\rightarrow \phi$ \vphantom{\LARGE h} &  \multicolumn{2}{c|}{-} &     \multicolumn{2}{c|}{-} &     \multicolumn{2}{c|}{-} & \multicolumn{2}{c|}{1.8$\cdot 10^{-2}$} & \multicolumn{2}{c|}{1.8$\cdot 10^{-4}$}      \\
gg$\rightarrow$ bb$\phi$ &  \multicolumn{2}{c|}{-} &     \multicolumn{2}{c|}{-} &     \multicolumn{2}{c|}{-} & \multicolumn{2}{c|}{2.0$\cdot 10^{-2}$} & \multicolumn{2}{c|}{2.6$\cdot 10^{-3}$}      \\
\hline
gg$\rightarrow$ H \vphantom{\LARGE h} &   \multicolumn{2}{c|}{1.7$\cdot 10^{-2}$} &  \multicolumn{2}{c|}{3.9$\cdot 10^{-4}$} &  \multicolumn{2}{c|}{1.1$\cdot 10^{-4}$} &    \multicolumn{2}{c|}{ -} &     \multicolumn{2}{c|}{-}       \\
qq$\rightarrow$ qqH &  \multicolumn{2}{c|}{8.6$\cdot 10^{-3}$} &  \multicolumn{2}{c|}{2.6$\cdot 10^{-3}$} &  \multicolumn{2}{c|}{5.2$\cdot 10^{-3}$} &     \multicolumn{2}{c|}{-} &     \multicolumn{2}{c|}{-}       \\
qq$\rightarrow\PH\ttbar$ or VH&  \multicolumn{2}{c|}{1.5$\cdot 10^{-2}$} &  \multicolumn{2}{c|}{3.3$\cdot 10^{-3}$} &  \multicolumn{2}{c|}{4.2$\cdot 10^{-5}$} &     \multicolumn{2}{c|}{-} &     \multicolumn{2}{c|}{-}       \\
\hline
\end{tabular}
\label{tab:cutflow-mutau}
\end{center}
\end{table*}

\begin{table*}[!htbp]
\begin{center}
\caption{Numbers of expected and observed events
  in the event categories as
  described in the text for the $\Pe\Pgm$+X channel. Also given are the expected signal yields and efficiencies for a MSSM
  Higgs boson with \mbox{$m_\mathrm{A}=120$\GeV} and $\tan\beta$ = 10, and for a SM Higgs boson with
  \mbox{$\mH=120$\GeV}. Combined statistical and systematic uncertainties on each estimate are reported.
  For the yield estimates for the Higgs signal the production cross sections for h and A, which have almost 
  degenerate masses, are taken into account.
  The quoted efficiencies do not include the branching fraction into $\Pgt\Pgt$.}
\begin{tabular}{|l|r@{$ \,\,\pm\,\, $}l|r@{$\,\,\pm\,\,$}l|r@{$\,\,\pm\,\,$}l|r@{$\,\,\pm\,\,$}l|r@{$\,\,\pm\,\,$}l|}
\cline{2-11}
 \multicolumn{1}{c}{ } & \multicolumn{6}{|c|}{SM} & \multicolumn{4}{c|}{MSSM} \\
 \hline
 Process & \multicolumn{2}{c|}{\emph{0/1-Jet}} & \multicolumn{2}{c|}{\emph{Boosted}} & \multicolumn{2}{c|}{\emph{VBF}} &  \multicolumn{2}{c|}{\emph{Non $b$-Tag}} & \multicolumn{2}{c|}{\emph{$b$-Tag}} \\
 \hline
 Z$\rightarrow \tau\tau$       & 11787 &   790 &    98 &    11 &    16 &     4 & 11718 &   797 &   112 &    11  \\
 Multijet and W+jets           &   483 &   145 &     9 &     3 &     2 &     1 &   474 &   147 &    15 &     5  \\
 $\ttbar$                &   427 &    41 &    70 &     8 &    14 &     3 &   161 &    15 &   289 &    35  \\
 Dibosons                      &   570 &    91 &    21 &     4 &   2.0 &   0.6 &   527 &    84 &    55 &    10  \\
 \hline
 \hline
 Total Background              & 13267 &   809 &   197 &    14 &    34 &     5 & 12881 &   815 &   471 &    38  \\
 \hline
 H$\rightarrow \tau\tau$       &    36 &     6 &   1.0 &   0.3 &   1.0 &   0.2 &   161 &    10 &    17 &   1.6  \\
 \hline
 Data                          & \multicolumn{2}{c|}{13152} & \multicolumn{2}{c|}{189} & \multicolumn{2}{c|}{26} & \multicolumn{2}{c|}{12761} & \multicolumn{2}{c|}{468}  \\
 \hline
 \multicolumn{6}{c}{ } \\
 \multicolumn{3}{l}{Signal Efficiency } &  \multicolumn{3}{c}{ } \\
 \hline
 gg$\rightarrow \phi$ \vphantom{\LARGE h}   & \multicolumn{2}{c|}{-} & \multicolumn{2}{c|}{-} & \multicolumn{2}{c|}{-} & \multicolumn{2}{c|}{6.4$\cdot 10^{-3}$} & \multicolumn{2}{c|}{9.4$\cdot 10^{-5}$} \\
 bb$\rightarrow$ bb$\phi$         & \multicolumn{2}{c|}{-} & \multicolumn{2}{c|}{-} & \multicolumn{2}{c|}{-} & \multicolumn{2}{c|}{5.8$\cdot 10^{-3}$} & \multicolumn{2}{c|}{9.8$\cdot 10^{-4}$} \\
 \hline
 gg$\rightarrow$ H \vphantom{\LARGE h}  & \multicolumn{2}{c|}{6.3$\cdot 10^{-3}$} & \multicolumn{2}{c|}{1.8$\cdot 10^{-4}$} & \multicolumn{2}{c|}{3.0$\cdot 10^{-5}$}& \multicolumn{2}{c|}{-} & \multicolumn{2}{c|}{-}  \\
 qq$\rightarrow$ qqH              & \multicolumn{2}{c|}{3.0$\cdot 10^{-3}$} & \multicolumn{2}{c|}{8.1$\cdot 10^{-4}$} & \multicolumn{2}{c|}{2.0$\cdot 10^{-3}$}& \multicolumn{2}{c|}{-} & \multicolumn{2}{c|}{-}  \\
 qq$\rightarrow\PH\ttbar$ or VH    & \multicolumn{2}{c|}{3.8$\cdot 10^{-3}$} & \multicolumn{2}{c|}{6.8$\cdot 10^{-4}$} & \multicolumn{2}{c|}{1.5$\cdot 10^{-6}$}& \multicolumn{2}{c|}{-} & \multicolumn{2}{c|}{-}  \\ \hline
\end{tabular}
\label{tab:cutflow-emu}
\end{center}
\end{table*}

\section{Tau-pair invariant mass reconstruction}

To distinguish the Higgs boson signal from the background, we
reconstruct the tau-pair mass using a maximum likelihood technique~\cite{CMS-PAPERS-HIG-10-002}.  The
algorithm estimates the original momentum components of the two taus
by maximizing a likelihood with respect to free parameters corresponding to the
missing neutrino momenta, subject to kinematic
constraints. Other terms in the likelihood take into account the
tau-decay phase space and the probability density in the tau
transverse momentum, parametrized as a function of the tau-pair mass.
This algorithm yields a tau-pair mass with a mean consistent with
the true value, and a distribution with a nearly Gaussian shape.
The standard deviation of the mass resolution is estimated to be 21\% at a Higgs boson mass of 130\GeV,
compared with 24\% for the (non-Gaussian) distribution of the
invariant mass spectrum reconstructed from the visible tau-decay products in the
inclusive selection. 
The resolution improves to 15\% in the b-Tag category in the MSSM analysis and in the boosted and VBF categories in the SM analysis where the Higgs
boson is produced with significant transverse momentum.

\section{Systematic uncertainties}

Various imperfectly known or simulated effects can alter
the shape and normalization of the invariant mass spectrum.
The main contributions to the normalization uncertainty include the uncertainty in the total
integrated luminosity (4.5\%)~\cite{CMS-LUMI}, jet energy scale (2--5\%
depending on $\eta$ and \pt), background normalization
(Tables~\ref{tab:cutflow-etau}--~\ref{tab:cutflow-emu}), $\cPZ$ boson
production cross section (2.5\%)~\cite{CMS-EWK-WZ}, lepton
identification and isolation efficiency (1.0\%), and trigger efficiency
(1.0\%). The tau-identification
efficiency uncertainty is estimated to be 6\% from an independent
study using a tag-and-probe technique~\cite{CMS-EWK-WZ}. The lepton identification and isolation
efficiencies are stable as a function of the number of additional
interactions in the bunch crossing in data and in Monte Carlo simulation.
The b-tagging efficiency carries an uncertainty of 10\%,
and the b-mistag rate is accurate to 30\%~\cite{BTV-11-001}.
Uncertainties that contribute to mass spectrum shape variations
include the tau (3\%), muon (1\%), and electron (1\% in the barrel
region, 2.5\% in the endcap region) energy scales.
The effect of the uncertainty on the $\MET$ scale, mainly due to
pile-up effects, is incorporated by varying the mass spectrum shape as
described in the next section.

The various production cross sections and branching fractions
for SM and MSSM Higgs bosons and corresponding uncertainties are taken from~\cite{LHCHiggsCrossSectionWorkingGroup:2011ti,Djouadi:1991tka,Dawson:1990zj,Spira:1995rr,Harlander:2002wh,Anastasiou:2002yz,Ravindran:2003um,Catani:2003zt,Aglietti:2004nj,Degrassi:2004mx,Actis:2008ug,Anastasiou:2008tj,deFlorian:2009hc,Baglio:2010ae,Ciccolini:2007jr,Ciccolini:2007ec,Arnold:2008rz,Brein:2003wg,Ciccolini:2003jy,hdecay2,Denner:2011mq,Botje:2011sn,Alekhin:2011sk,Lai:2010vv,Martin:2009iq,Ball:2011mu,Djouadi:1991tka,Aglietti:2004nj,Degrassi:2004mx,Baglio:2010ae}. Theoretical uncertainties on the Higgs production cross section are included in the SM and the MSSM search. For the SM signal, these uncertainties range from 12-30\% for gluon fusion, depending on the event category, and 10\% for VBF production. The uncertainty for the MSSM signal depends on $\tan \beta$ and $m_{A}$ and ranges from 20-25\%.

\section{Maximum likelihood fit}

To search for the presence of a Higgs boson signal in the selected
events, we perform a binned maximum likelihood fit to the tau-pair
invariant-mass spectrum, $\Mfit$.
The fit is performed jointly across the three SM and two MSSM event categories, but independently in the two cases.

Systematic uncertainties are represented by nuisance parameters in the fitting process.
We assume log-normal priors for
normalization parameters, and Gaussian priors for mass-spectrum shape
uncertainties. The uncertainties that affect the shape of the mass
spectrum, mainly those corresponding to the energy scales, are
represented by nuisance parameters whose variation results in a
continuous perturbation of the spectrum shape~\cite{Conway-PhyStat}.

\section{Results}

\begin{figure}[htbp]
\begin{center}
\includegraphics[width=0.48\textwidth]{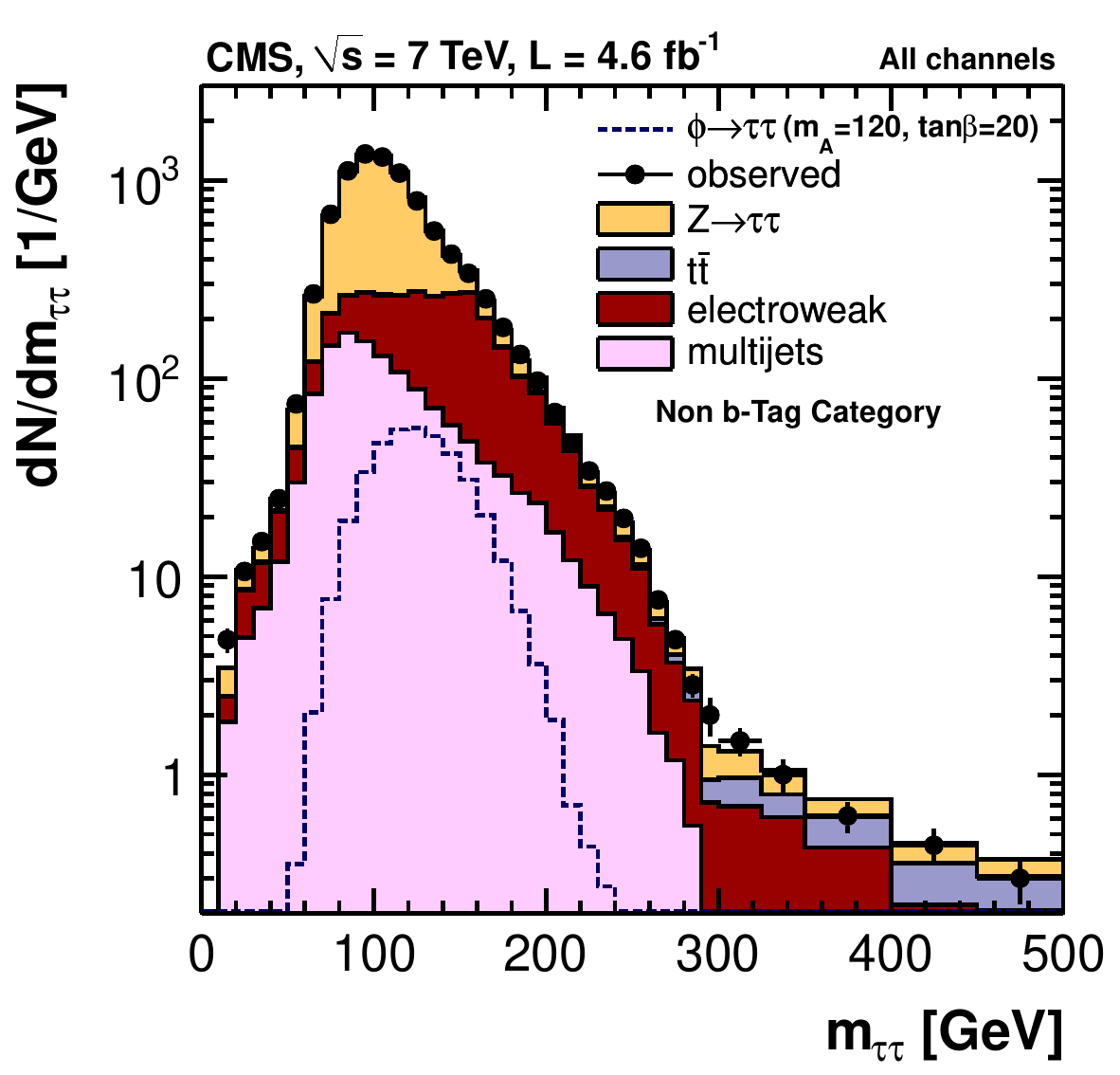}
\includegraphics[width=0.48\textwidth]{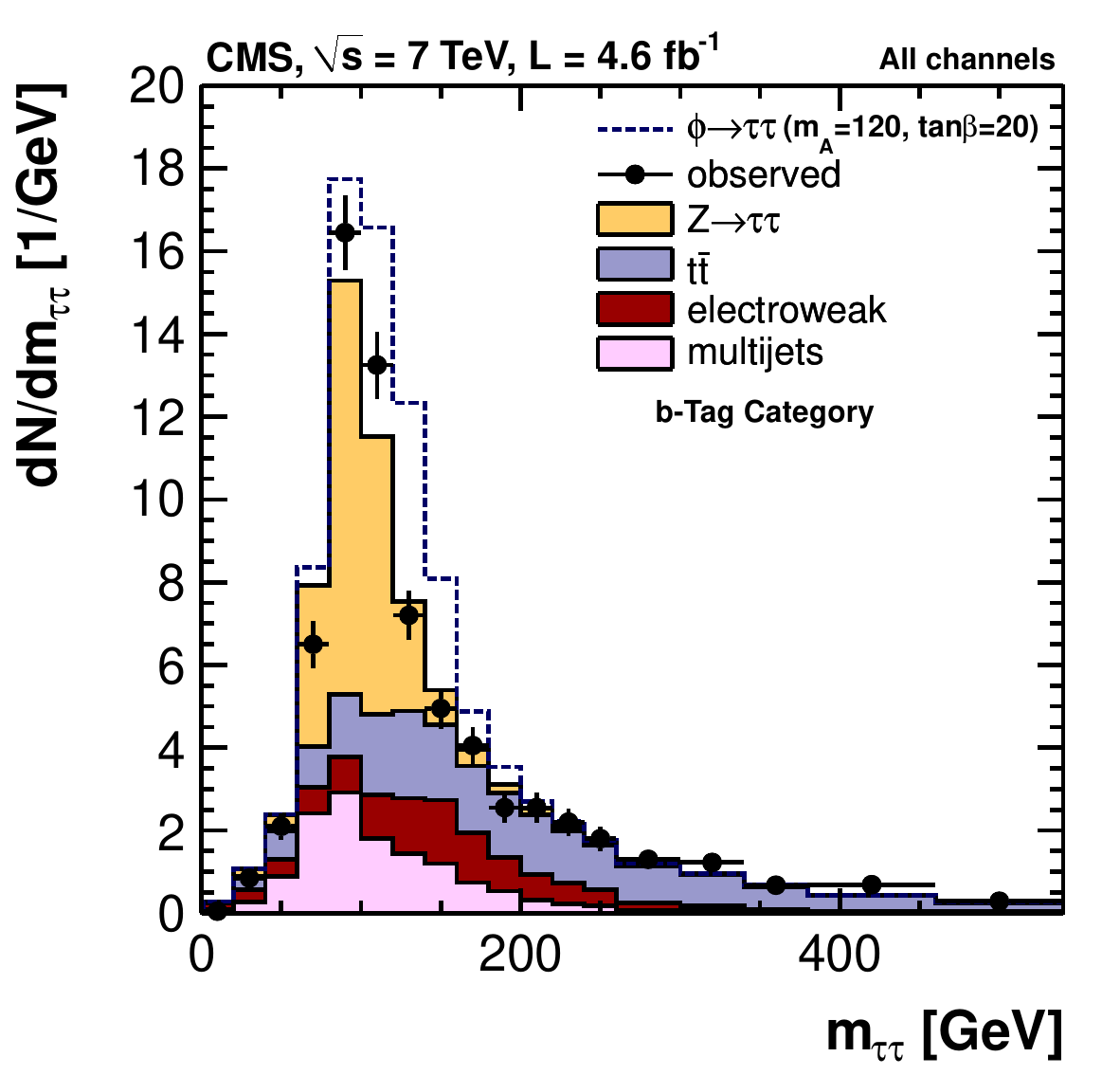}
\end{center}
\caption{Distribution of the tau-pair invariant mass, $\Mfit$, in the MSSM Higgs boson search
   categories: Non b-Tag category (left), b-Tag category (right). The background labelled 'electroweak' combines the contribution from W+jets, Z$\rightarrow ll$, and diboson processes.}
\label{fig:mfit_mssm}
\end{figure}

\begin{figure*}[!htbp]
\begin{center}
\includegraphics[width=0.48\textwidth]{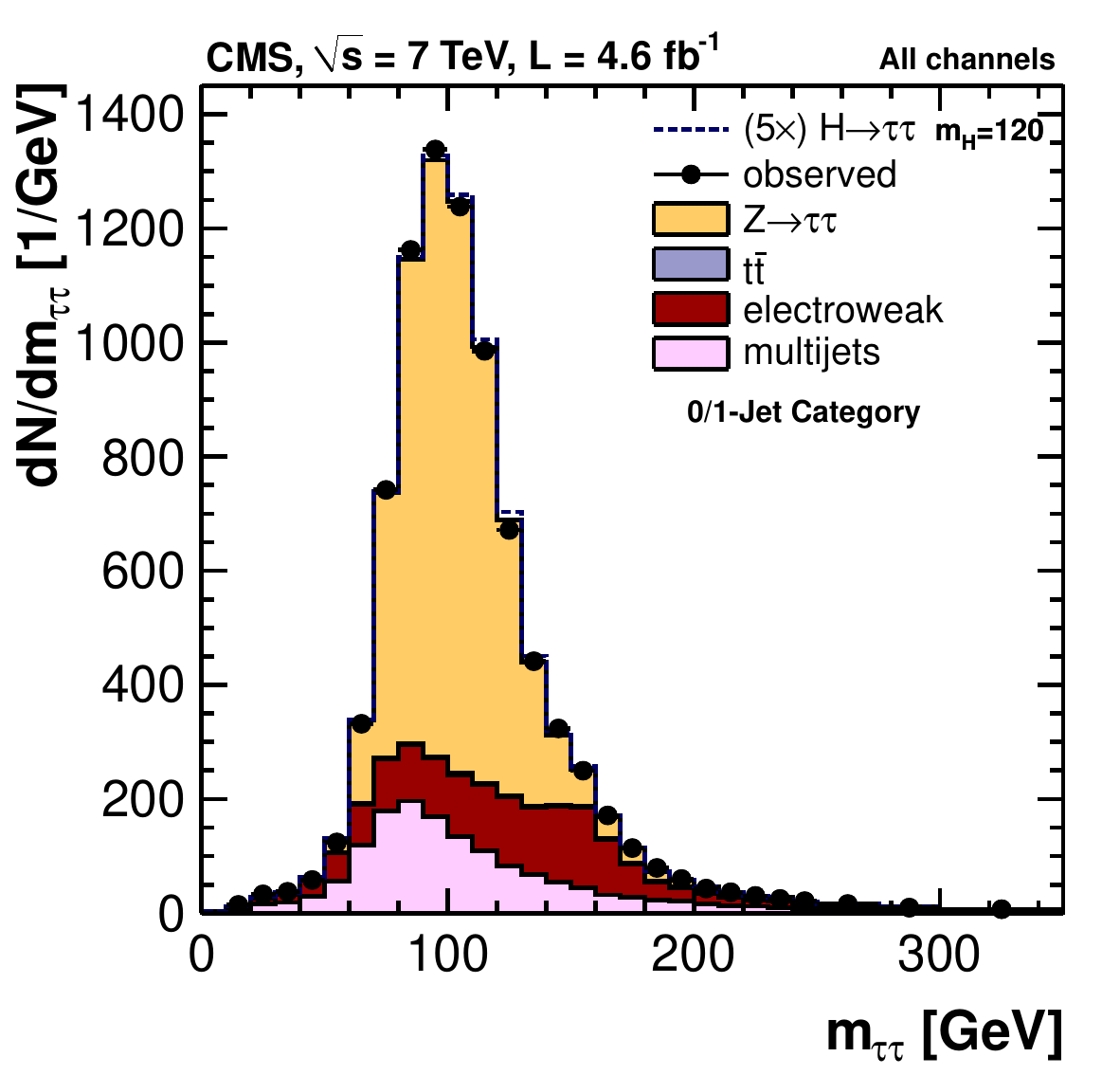}
\includegraphics[width=0.48\textwidth]{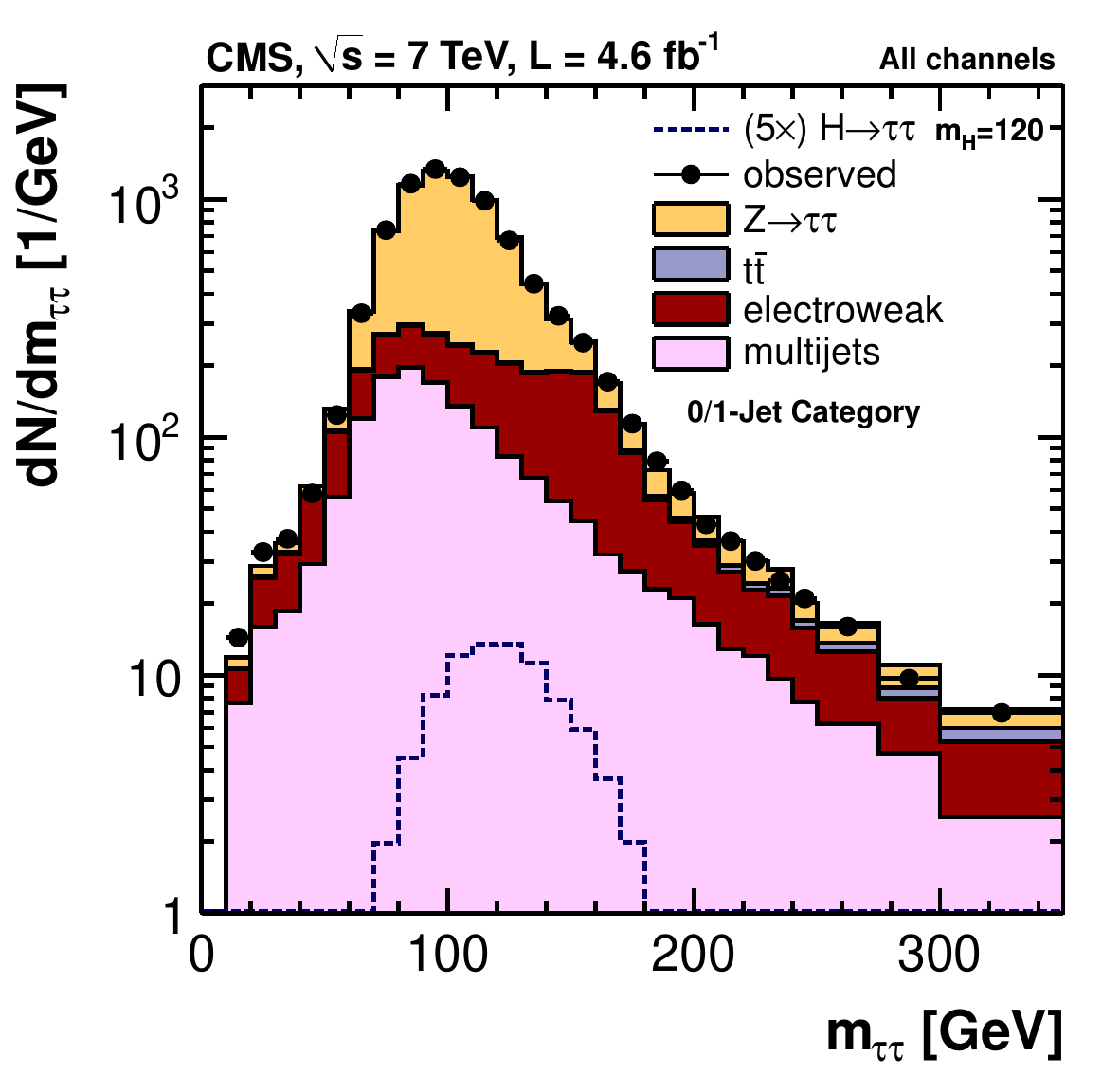}
\includegraphics[width=0.48\textwidth]{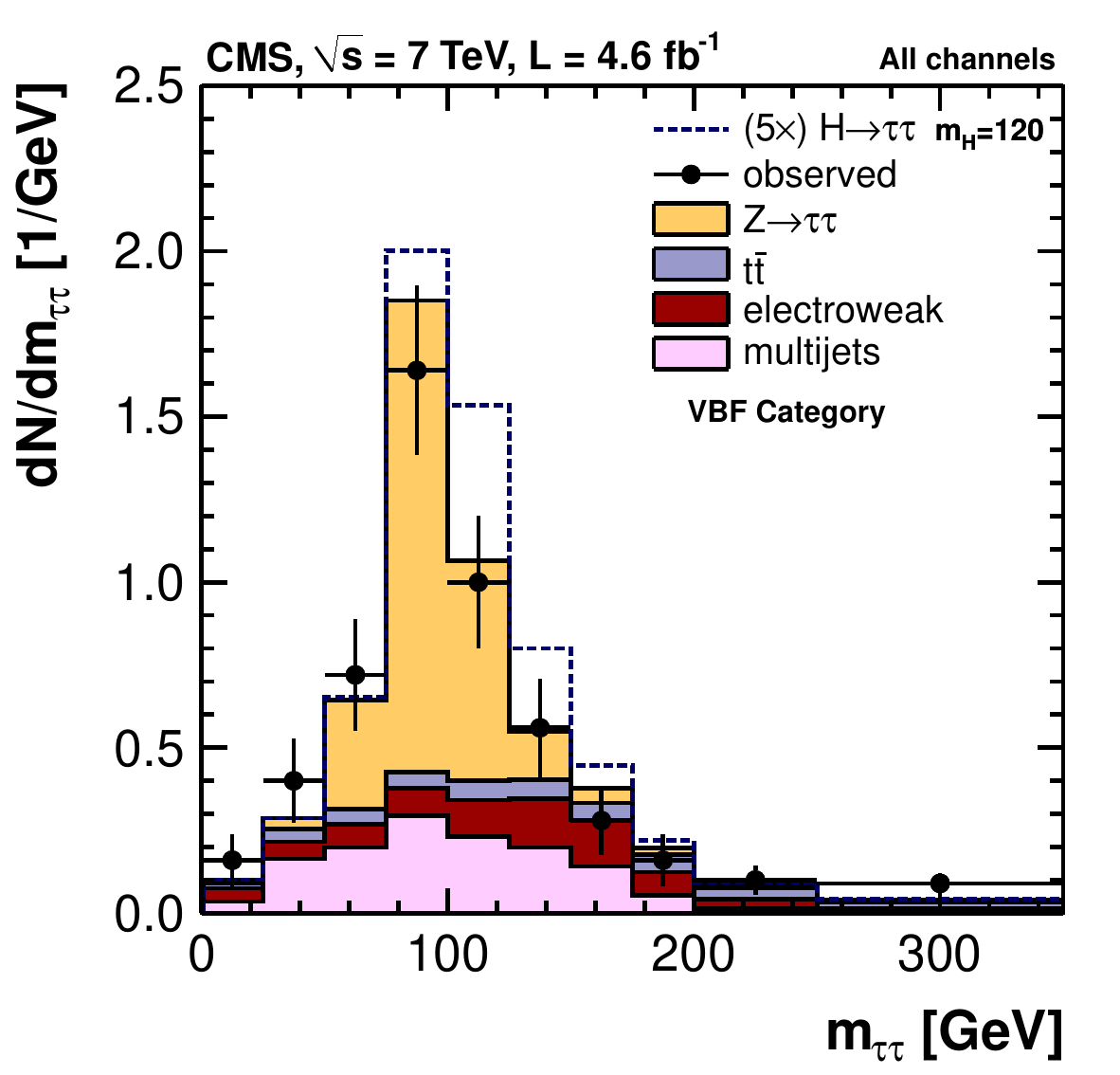}
\includegraphics[width=0.48\textwidth]{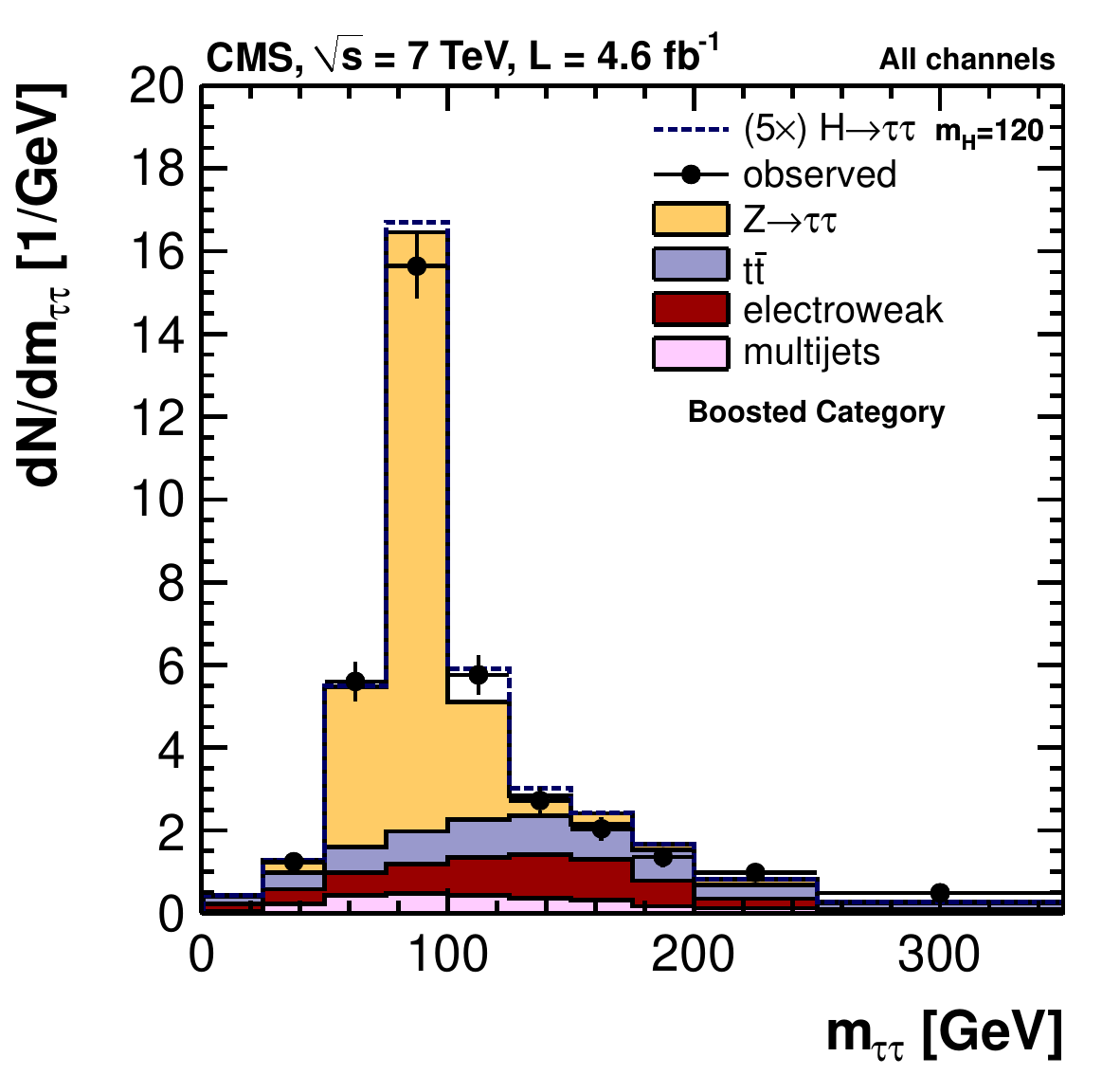}
\end{center}
\caption{Distribution of the tau-pair invariant mass, $\Mfit$, in the SM Higgs boson search
  categories: 0/1 Jet (top row, linear and log vertical scale),
  VBF (lower left), and Boosted (lower right). The background labelled 'electroweak' combines the contribution from W+jets, Z$\rightarrow ll$, and diboson processes.}
\label{fig:mfit_sm}
\end{figure*}

Figures~\ref{fig:mfit_mssm} and \ref{fig:mfit_sm} show for the SM and MSSM, respectively, the
distributions of the tau-pair mass $\Mfit$ summed over the three search
channels, for each category, compared with the background prediction.
The background mass distributions show the results of the fit using the background-only hypothesis.

The invariant mass spectra for both the MSSM and SM categories show no
evidence for the presence of a Higgs boson signal, and we therefore set 95\% CL upper bounds on the Higgs boson cross section times the branching fraction into a tau pair. For calculations of exclusion limits, we
use the modified frequentist construction CL$_s$~\cite{Junk,Read,LHC-HCG}.
Theoretical uncertainties on the Higgs boson production cross sections are taken into account
as systematic uncertainties in the limit calculations.

\subsection{Limits on MSSM Higgs boson production}

For the $m_\mathrm{h}^{\rm max}$ benchmark scenario as described above we set a 95\% CL upper limit on $\tan\beta$ as a function of the pseudoscalar Higgs boson mass $m_\mathrm{A}$ from the observed di-tau mass distributions in the b-Tag and non b-Tag event categories. Signal contributions from h, H and A production are considered. The mass values of h and H, as well as the ratio between the gluon fusion process and the associated production with b quarks, depend on the value of $\tan\beta$. 
To account for this, we perform a scan of $\tan\beta$ for each mass hypothesis, using the Higgs boson cross sections as a function of $\tan\beta$ as reported by the LHC Cross Section Working Group~\cite{LHCHiggsCrossSectionWorkingGroup:2011ti}. For the gluon-fusion process these cross sections have been obtained from the {\sc GGH@NNLO}~\cite{Harlander:2002wh,Harlander:2002vv,Anastasiou:2002wq} and {\sc HIGLU}~\cite{Spira:1995mt} programs. For the $\cPqb\cPaqb\to\phi$ process, the four-flavor calculation~\cite{Dittmaier:2003ej,Dawson:2003kb} and the five-flavor calculation as implemented in the {\sc BBH@NNLO}~\cite{Harlander:2003ai} program have been combined using the Santander scheme~\cite{Santander}. Rescaling of the corresponding Yukawa couplings by the MSSM factors calculated with {\sc FeynHiggs}~\cite{Heinemeyer:1998yj,Heinemeyer:1998np,Degrassi:2002fi} has been applied.

\begin{figure}[hbtp]
  \begin{center}
    \includegraphics[width=\cmsFigWidth]{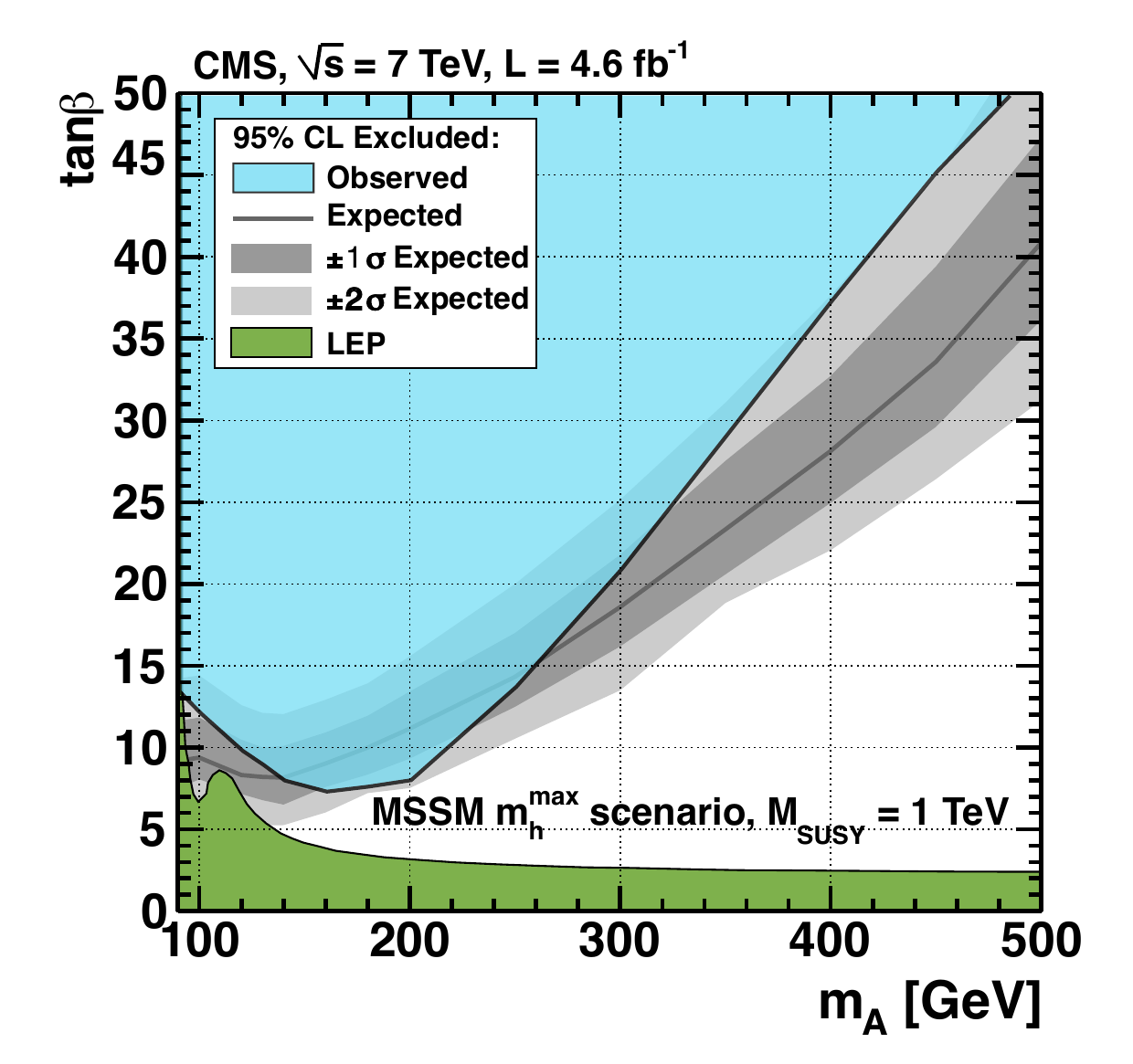} 
    \caption{Region in the parameter space of $\tan\beta$ versus $m_\mathrm{A}$ excluded
             at 95\% CL in the context of the MSSM $m^\mathrm{max}_\mathrm{h}$ scenario.
             The expected one- and two-standard-deviation ranges and the observed 95\%
             CL upper limits are shown together with the observed excluded region.}
    \label{tanbeta-ma}
  \end{center}
\end{figure}

Figure~\ref{tanbeta-ma} also shows the region excluded by the LEP experiments~\cite{LEP2-MSSM}.
The results reported in this Letter considerably extend the exclusion region of the MSSM parameter space and
supersede limits reported by CMS using a smaller data sample collected in 2010~\cite{CMS-PAPERS-HIG-10-002}.

\begin{table*}[htbp]
  \begin{center}
    \caption{Expected range and observed 95\% CL upper limits for
             $\tan\beta$ as a function of $m_\mathrm{A}$, for the MSSM search.}
\begin{tabular}{|c|c|c|c|c|c|c|}
\cline{2-6}
\multicolumn{1}{c}{ MSSM Higgs }      & \multicolumn{5}{|c|}{Expected $\tan\beta$ limit}   &  \multicolumn{1}{c}{}\\
\hline
  $m_\mathrm{A}$ [GeV] &$-2\sigma$  &   $-1\sigma$ &        Median &    $+1\sigma$ &  $+2\sigma$ & Obs. $\tan\beta$ limit \\ \hline
    90 &  5.19 &  7.01 &   8.37 &  10.6 &  12.8 &   12.2  \\  \hline
   100 &  6.49 &  7.45 &   8.78 &  10.8 &  13.4 &   11.8  \\  \hline
   120 &  4.50 &  6.47 &   8.09 &  9.89 &  12.0 &   9.84  \\  \hline
   130 &  5.37 &  6.71 &   7.85 &  9.69 &  11.5 &   9.03  \\  \hline
   140 &  5.62 &  6.63 &   7.90 &  9.69 &  11.6 &   8.03  \\  \hline
   160 &  5.57 &  6.99 &   8.51 &  10.4 &  12.5 &   7.11  \\  \hline
   180 &  6.75 &  8.14 &   9.53 &  11.3 &  13.8 &   7.50  \\  \hline
   200 &  7.84 &  9.12 &   10.5 &  12.8 &  15.0 &   8.46  \\  \hline
   250 &  10.3 &  12.3 &   13.9 &  16.8 &  19.4 &   13.8  \\  \hline
   300 &  13.5 &  15.7 &   18.4 &  21.4 &  24.5 &   20.9  \\  \hline
   350 &  17.7 &  20.1 &   23.0 &  26.9 &  31.1 &   29.1  \\  \hline
   400 &  21.9 &  24.3 &   27.9 &  32.4 &  37.3 &   37.3  \\  \hline
   450 &  25.0 &  29.2 &   33.3 &  38.8 &  44.7 &   45.2  \\  \hline
   500 &  30.3 &  35.7 &   40.5 &  47.1 &  55.0 &   51.9  \\  \hline
 \end{tabular}
    \label{tab-limits}
  \end{center}
\end{table*}

\subsection{Limits on SM Higgs boson production}

\begin{figure}[hbtp]
  \begin{center}
    \includegraphics[width=\cmsFigWidth]{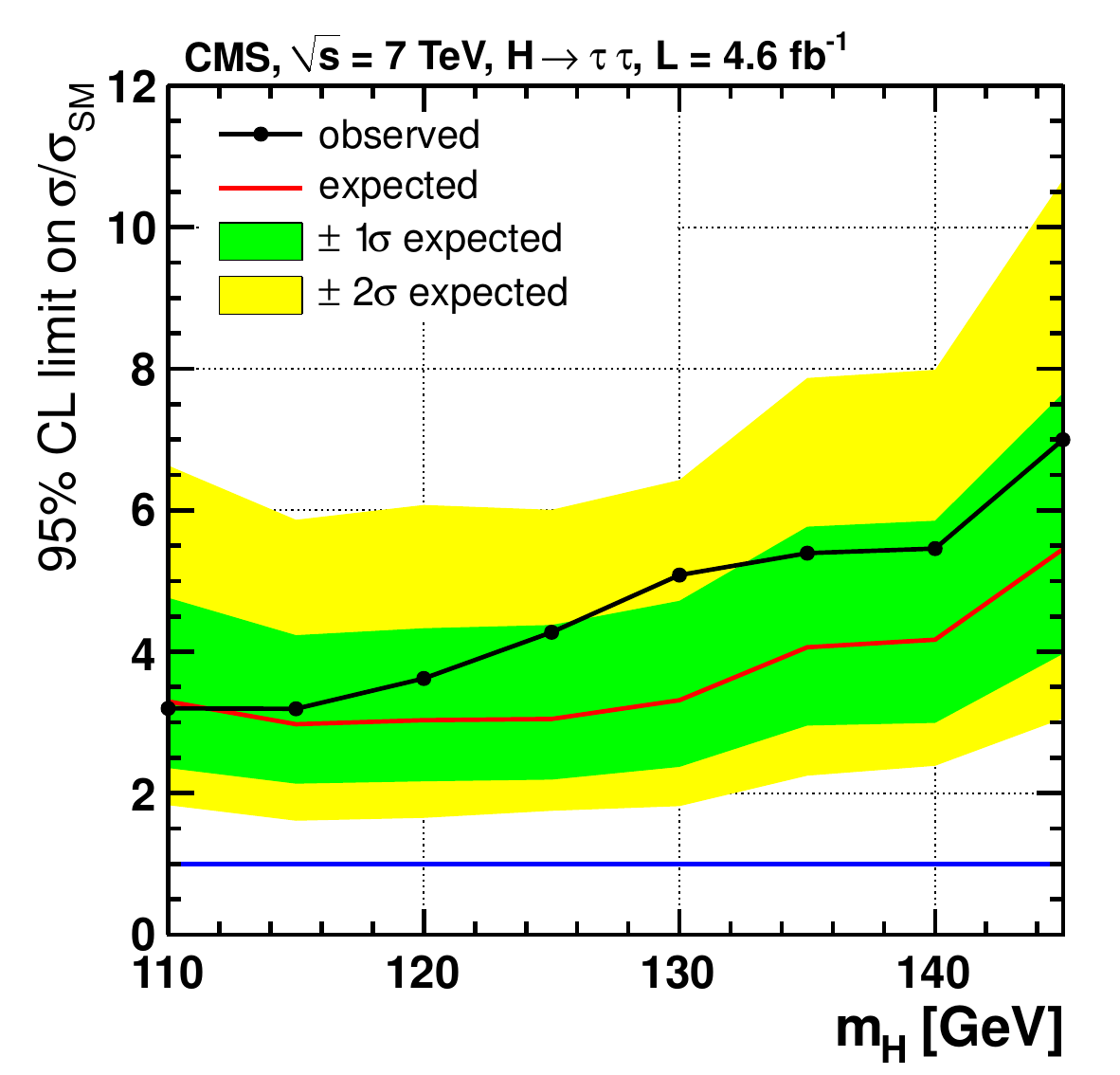}
    \caption{The expected one- and two-standard-deviation ranges are shown together with the observed 95\%
       CL upper limits on the cross section, normalized to the SM expectation for Higgs boson production, as a function of $\mH$.}
    \label{sm-limit}
  \end{center}
\end{figure}

The 0/1 Jet, VBF and Boosted categories are used to set a 95\% CL upper limit
on the product of the Higgs boson production cross section and the $\PH \to \tau \tau$ branching fraction,
$\sigma_{\PH} \times \mathrm{BR}(\PH \to \Pgt\Pgt)$, with respect to the
SM Higgs expectation, $\sigma/\sigma_{\text{SM}}$. Figure~\ref{sm-limit} shows the observed and
the mean expected 95\% CL upper limits for Higgs boson mass hypotheses
ranging from 110 to 145\GeV. The bands represent the one- and two-standard-deviation
probability intervals around the expected limit. Table~\ref{tab-sm-limits} shows the results for selected mass
values. We set a 95\% upper limit on $\sigma/\sigma_{\text{SM}}$ in the range of 3--7.

\begin{table*}[htbp]
  \begin{center}
    \caption{Expected range and observed 95\% CL upper limits on the cross section, divided by the expected SM Higgs cross section as a function of $\mH$, for the SM search.}
\begin{tabular}{|c|c|c|c|c|c|c|}
\cline{2-6}
\multicolumn{1}{c}{ SM Higgs }      & \multicolumn{5}{|c|}{Expected limit}     &  \multicolumn{1}{c}{ }  \\
\hline
           $\mH$ [GeV] &     $-2\sigma$  &      $-1\sigma$ &          Median &      $+1\sigma$ &      $+2\sigma$ & Obs. limit  \\
\hline
              110 &            1.83 &            2.36 &            3.30 &            4.76 &            6.63 &            3.20  \\
\hline
              115 &            1.61 &            2.13 &            2.97 &            4.23 &            5.86 &            3.19  \\
\hline
              120 &            1.65 &            2.17 &            3.03 &            4.33 &            6.07 &            3.62  \\
\hline
              125 &            1.75 &            2.19 &            3.05 &            4.38 &            6.01 &            4.27  \\
\hline
              130 &            1.82 &            2.37 &            3.31 &            4.72 &            6.43 &            5.08  \\
\hline
              135 &            2.25 &            2.96 &            4.06 &            5.77 &            7.87 &            5.39  \\
\hline
              140 &            2.39 &            2.99 &            4.17 &            5.85 &            7.99 &            5.46  \\
\hline
              145 &            3.06 &            3.97 &            5.45 &            7.65 &            10.7 &            7.00  \\
\hline
\end{tabular}
    \label{tab-sm-limits}
  \end{center}
\end{table*}

\section{Summary}
We have reported a search for SM and neutral MSSM Higgs bosons, using a sample of CMS data from proton-proton collisions at a center-of-mass energy of 7\TeV at the LHC, corresponding to an integrated luminosity of 4.6\fbinv.  The tau-pair decay mode in final states with one $\Pe$ or $\Pgm$ plus a hadronic decay of a tau, and the $\Pe\Pgm$ final state are used. The observed tau-pair mass spectra reveal no evidence for neutral Higgs boson production. In the SM case we determine a 95\% CL upper limit in the mass range of 110--145\GeV on the Higgs boson production cross section. We exclude a Higgs boson with $\mH=$115\GeV with a production cross section 3.2 times of that predicted by the standard model. In the MSSM case, we determine a 95\%~CL upper bound on the value of $\tan\beta$ as a function of $m_\mathrm{A}$, for the $m_\mathrm{h}^\mathrm{max}$ scenario. This search excludes a previously unexplored region reaching as low as $\tan\beta = 7.1$ at $m_\mathrm{A} = 160$\GeV.

\section*{Acknowledgements}
We congratulate our colleagues in the CERN accelerator departments for the excellent performance of the LHC machine. We thank the technical and administrative staff at CERN and other CMS institutes, and acknowledge support from: FMSR (Austria); FNRS and FWO (Belgium); CNPq, CAPES, FAPERJ, and FAPESP (Brazil); MES (Bulgaria); CERN; CAS, MoST, and NSFC (China); COLCIENCIAS (Colombia); MSES (Croatia); RPF (Cyprus); MoER, SF0690030s09 and ERDF (Estonia); Academy of Finland, MEC, and HIP (Finland); CEA and CNRS/IN2P3 (France); BMBF, DFG, and HGF (Germany); GSRT (Greece); OTKA and NKTH (Hungary); DAE and DST (India); IPM (Iran); SFI (Ireland); INFN (Italy); NRF and WCU (Korea); LAS (Lithuania); CINVESTAV, CONACYT, SEP, and UASLP-FAI (Mexico); MSI (New Zealand); PAEC (Pakistan); MSHE and NSC (Poland); FCT (Portugal); JINR (Armenia, Belarus, Georgia, Ukraine, Uzbekistan); MON, RosAtom, RAS and RFBR (Russia); MSTD (Serbia); MICINN and CPAN (Spain); Swiss Funding Agencies (Switzerland); NSC (Taipei); TUBITAK and TAEK (Turkey); STFC (United Kingdom); DOE and NSF (USA). Individuals have received support from the Marie-Curie program and the European Research Council (European Union); the Leventis Foundation; the A. P. Sloan Foundation; the Alexander von Humboldt Foundation; the Belgian Federal Science Policy Office; the Fonds pour la Formation \`a la Recherche dans l'Industrie et dans l'Agriculture (FRIA-Belgium); the Agentschap voor Innovatie door Wetenschap en Technologie (IWT-Belgium); the Council of Science and Industrial Research, India; and the HOMING PLUS program of Foundation for Polish Science, cofinanced from European Union, Regional Development Fund.

\bibliography{auto_generated}   

\cleardoublepage \appendix\section{The CMS Collaboration \label{app:collab}}\begin{sloppypar}\hyphenpenalty=5000\widowpenalty=500\clubpenalty=5000\input{HIG-11-029-authorlist.tex}\end{sloppypar}
\end{document}

%% file: HIG-11-029-authorlist.tex
\textbf{Yerevan Physics Institute,  Yerevan,  Armenia}\\*[0pt]
S.~Chatrchyan, V.~Khachatryan, A.M.~Sirunyan, A.~Tumasyan
\vskip\cmsinstskip
\textbf{Institut f\"{u}r Hochenergiephysik der OeAW,  Wien,  Austria}\\*[0pt]
W.~Adam, T.~Bergauer, M.~Dragicevic, J.~Er\"{o}, C.~Fabjan, M.~Friedl, R.~Fr\"{u}hwirth, V.M.~Ghete, J.~Hammer\cmsAuthorMark{1}, M.~Hoch, N.~H\"{o}rmann, J.~Hrubec, M.~Jeitler, W.~Kiesenhofer, M.~Krammer, D.~Liko, I.~Mikulec, M.~Pernicka$^{\textrm{\dag}}$, B.~Rahbaran, C.~Rohringer, H.~Rohringer, R.~Sch\"{o}fbeck, J.~Strauss, A.~Taurok, F.~Teischinger, P.~Wagner, W.~Waltenberger, G.~Walzel, E.~Widl, C.-E.~Wulz
% ...
\vskip\cmsinstskip
\dag:~Deceased\\
1:~~Also at CERN, European Organization for Nuclear Research, Geneva, Switzerland\\
% ...
